\documentclass[10pt,letterpaper]{article}
\usepackage[top=0.85in,left=2.75in,footskip=0.75in]{geometry}

\usepackage{amsmath,amssymb}

\usepackage{changepage}

\usepackage[utf8x]{inputenc}

\usepackage{textcomp,marvosym}

\usepackage{cite}

\usepackage{nameref,hyperref}

\usepackage[right]{lineno}

\usepackage{microtype}
\DisableLigatures[f]{encoding = *, family = * }

\usepackage[table]{xcolor}

\usepackage{array}

\usepackage{enumerate}

\newcolumntype{+}{!{\vrule width 2pt}}

\newlength\savedwidth

\usepackage{color}

\raggedright
\usepackage[aboveskip=1pt,labelfont=bf,labelsep=period,justification=raggedright,singlelinecheck=off]{caption}

\makeatletter
\renewcommand{\@biblabel}[1]{\quad#1.}
\makeatother

\date{}

\usepackage{lastpage,fancyhdr,graphicx}
\usepackage{epstopdf}
\fancyheadoffset[L]{2.25in}
\fancyfootoffset[L]{2.25in}
\newcommand{\withurl}[2]{{#1}\footnote{\texttt{#2}}}
\newcommand{\rulemajor}[1]{\section{#1}}
\newcommand{\ruleminor}[1]{\textbf{#1}}
\newcommand{\reviewed}[1]{{\color{black}#1}}
\begin{document}
\vspace*{0.2in}

\begin{flushleft}
{\Large
\textbf\newline{Ten Simple Rules for Making Research Software More Robust}
}
\newline
\\
{Morgan~Taschuk}\textsuperscript{1,\ddag *},
{Greg~Wilson}\textsuperscript{2,\ddag}
\\
\textbf{1} Ontario Institute for Cancer Research / morgan.taschuk@oicr.on.ca
\\
\textbf{2} Software Carpentry Foundation / gvwilson@software-carpentry.org
\\
\bigskip
{\ddag} These authors contributed equally to this work.
\\
* Corresponding author.
\end{flushleft}

\section*{Abstract}

Software produced for research,
published and otherwise,
suffers from a number of common problems
that make it difficult or impossible to run outside the original institution,
or even off the primary developer's computer.
We present ten simple rules to make such software robust enough to be run by anyone,
anywhere,
and thereby delight your users and collaborators.

\section*{Author Summary}

Many researchers have found out the hard way that there's a world of difference
between ``works for me on my machine'' and ``works for other people on theirs''.
Many common challenges can be avoided by following a few simple rules; doing so
not only improves reproducibility, but can accelerate research.

% \linenumbers

\section*{Introduction}

Scientific software is typically developed and used by a single person,
usually a graduate student or postdoc~\cite{prins2015}.
It may produce the intended results in their hands,
but what happens when someone else wants to run it? Everyone with
a few years of experience feels a bit nervous when told to use
another person's code to analyze their data:
it will often be undocumented,
work in unexpected ways (if it works at all),
rely on nonexistent paths or resources,
be tuned for a single dataset,
or simply be an older version than was used in published papers.
The potential new user is then faced with two unpalatable options:
hack the existing code to make it work, or start over.

Being unable to
replicate results is so common that one publication refers to it as ``a rite of
passage''~\cite{baker2016}.
The root cause of this problem is that most research software
\reviewed{ is essentially a prototype, and therefore is not
\emph{robust}. The lack of
robustness in published, distributed software
leads to duplicated efforts with little practical benefit,}
which slows the pace of research~\cite{prabhu2011,lawlor2015}.
Bioinformatics software repositories \cite{ison2016,brazas2012} catalogue dozens to
hundreds of tools that perform similar tasks:
for example,
in 2016 the Bioinformatics Links Directory included 84 different multiple sequence aligners, 141 tools
to analyze transcript expression, and 182 pathway and interaction resources.
Some of these tools are legitimate efforts to improve the state-of-the-art, but
often they are difficult to install and run~\cite{stajich2002,Seemann2013}, and are effectively abandoned
after publication~\cite{nekrutenko2012}.

This problem is not unique to bioinformatics, or even to computing~\cite{baker2016}.
Best practices in software engineering specifically aim to increase software
robustness. However, most bioinformaticians learn what they know about software development
on the job or otherwise informally~\cite{prins2015,atwood2015}.
Existing training programs and initiatives rarely have the time to cover software engineering
in depth, especially since the field is so broad and developing so rapidly~\cite{atwood2015,lawlor2015}.
In addition, making software robust is not directly rewarded
in science, and funding is difficult to come by~\cite{prins2015}. Some proposed
solutions to this problem include restructuring educational programs,
hiring dedicated software engineers~\cite{lawlor2015,sanders2008},
partnering with private sector or grassroots organizations~\cite{prins2015,ison2016},
or using specific technical tools like containerization or cloud
computing~\cite{afgan2016,howe2012}. Each of these requires time and, in some
cases, institutional change.

The good news is,
you don't need to be a professionally-trained programmer to write robust software.
In fact,
some of the best, most reliable pieces of software in many scientific
communities are written by researchers~\cite{prabhu2011,sanders2008}
who have adopted strong software
engineering approaches, have high standards of reproducibility, use good testing
practices, and foster 
\reviewed{strong user
bases through constantly evolving, clearly documented, useful, and useable software.}
In the bioinformatics community, Bioconductor and Galaxy follow this
path~\cite{gentleman2004,afgan2016}.
\reviewed{Not all scientific software needs to be robust~\cite{varoquaux2015}, but if you
publish a paper about your software, it should, at minimum, satisfy these
rules.}

So what \emph{is} ``robust'' software?
\reviewed{We implied above that it is software that works for people other than the original author,
on machines other than its creator's.
More specifically, we mean that:

\begin{itemize}
\item
  it can be installed on more than one computer with relative ease,
\item
  it works consistently as advertised, and
\item
  it can be integrated with other tools.
\end{itemize}

Our rules are generic and can be applied to all languages, libraries, packages,
documentation styles, and operating systems for both closed-source and open-source software. 
They are also necessary steps toward making computational research replicable and reproducible:
after all,
if your tools and libraries cannot be run by others,
they cannot be used to verify your results or as a stepping stone for future work~\cite{brown2013}.}

\reviewed{\rulemajor{Rule 1: Use version control.}

Version control is essential to sustainable software development
\cite{wilson2014,wilson2016}.
In particular,
developers will struggle to understand what they have actually built,
what it actually does,
and what they have actually released
without some mechanical way to keep track of changes.
They should therefore
\ruleminor{put everything created manually into version control as soon as it is created},
including programs, original field observations, and the source files for papers.
Files that can be regenerated at need,
such as the binaries for compiled programs or intermediate files generated during data analysis,
should not be versioned;
instead,
it is often more sensible to use an archiving system for them,
and store the metadata describing their contents in version control instead~\cite{noble2009}.

If you are new to version control,
it is simplest to treat it as ``a better \withurl{Dropbox}{http://dropbox.com}''
(or, if you are of a certain age, a better FTP),
and to use it simply to synchronize files between multiple developers and machines~\cite{blischak2016}.
Once you are comfortable working that way,
you should \ruleminor{use a feature branch workflow}:
designate one parallel copy (or ``branch'') of the repository as the master,
and create a new branch from it each time you want to fix a bug or add a new feature.
This allows work on independent changes to proceed in isolation;
once the work has been completed and tested,
it can be merged into the master branch for release.}

\reviewed{\rulemajor{Rule 2: Document your code and usage}

How to write high quality documentation has been described elsewhere~\cite{karimzadeh2016}
and so here we only cover two minimal types: the README and usage.
The README is usually available even before the software is installed,
exists
to get a new user started, and points them towards more
help. Usage is a terse, informative command-line help message that
guides the user in the correct use of the software.

Numerous guidelines exist on how to \ruleminor{write a good
README file}~\cite{Johnson1997,gnustandards}.
At a minimum, your README should:

\begin{enumerate}

\item \textit{Explain what the software does}.
There's nothing more frustrating
than downloading and installing something only to
find out that it doesn't do what you thought it did.

\item \textit{List required dependencies}.
We address dependencies in more detail in Rule~5.

\item \textit{Provide compilation or installation instructions}.

\item \textit{List all input and output files},
even those considered self-explanatory.
Link to specifications for standard formats
and list the required fields and acceptable values in other files.
If there is no rigorous format, explain what the sections mean.
These files are often full of valuable information that can be
mined for the user's specific purpose.

\reviewed{\item \textit{List a few example commands} to get a user started quickly.}

\item \textit{State attributions and licensing}. Attributions are how you credit
your contributors; licenses dictate how others may use and
need to credit your work.
\end{enumerate}

The program should also
\ruleminor{print usage information} when launching from the command line.
Usage provides the first line of help for both new and experienced users.
Terseness is important: usage that extends for multiple screens is
difficult to read or refer to on the fly.

Almost all command-line applications use a combination of
POSIX~\cite{posix2016} and GNU~\cite{gnustandards} standards for usage.
More standard command-line behaviours are detailed in \cite{Seemann2013}.
Your software's usage should:

\begin{enumerate}

\item \textit{Describe the syntax for running the program}, including the
  name of the program, the relative location of optional
  and required flags, other arguments, and values for execution.

\item \textit{Give a short description} to remind users of the software's primary function.

\item \textit{List the most commonly used arguments}, a description of each, and
    the default values.

\item \textit{State where to find more information}.

\end{enumerate}

Usage should be
printed to standard output so that it can be
combined with other bash utilities like \texttt{grep},
and it should finish with an appropiate exit code.

Documentation beyond the README and usage is up to the developer's discretion.
We think it is very important for developers to document their work, but our
experience is that people are unlikely do it during normal development
However, it
is worth noting that software that is widely used and contributed to has and
enforces the need for good documentation~\cite{gentleman2004}.}

\rulemajor{Rule 3: Make common operations easy to control.}

Being able to change parameters on the fly to determine if and how
they change the results is important as your software gains more users,
as it facilitates exploratory analysis and parameter sweeping.
Programs should therefore
\ruleminor{allow the most commonly changed parameters to be configured from the command line}.

Users will want to change some values more often than others.
Since parameters are software-specific, the appropriate 'tunable' ones cannot be detailed here,
but a short list includes input and reference files and directories,
output files and directories,
filtering parameters,
random number generation seeds,
and
alternatives such as compressing results,
use a variant algorithm,
or verbose output.

\ruleminor{Check that all input values are in a reasonable range at startup}.
Few things are as annoying as having a program announce after running for two
hours that it isn't going to save its results because the requested directory
doesn't exist.

To make programs even easier to use,
\ruleminor{choose reasonable defaults where they exist}
and \ruleminor{set no defaults at all when there aren't any reasonable ones}.
You can set reasonable default values
as long as any command line arguments
override those values.

Changeable values should \emph{never} be hard-coded:
if users have to edit your software in order to run it,
you have done something wrong.
Changeable but infrequently-changed values should therefore be stored in configuration files.
These can be in a standard location,
e.g. \texttt{.packagerc} in the user's home directory,
or provided on the command line as an additional argument.
Configuration files are often created during installation
to set up such things as server names,
network drives,
and other defaults for your lab or institution.

\rulemajor{Rule 4: Version your releases.}

Software evolves over time, with developers adding or removing features as need
dictates. Making official releases stamps a particular set of features with a
project-specific identifier so that version can be retrieved for later use. For
example, if a paper is published, the software should be released at the same
time so that the results can be reproduced.

Most software has a version number composed of a decimal number that
increments as new versions are released.
There are many different ways
to construct and interpret this number, but most importantly for us, a
particular software version run with the same parameters should give
identical results no matter when it's run. Results include both correct
output as well as any errors.
\ruleminor{Increment your version number every time you release your software to
other people}.

\withurl{Semantic versioning}{http://semver.org/} is one of the most common
types of versioning for open-source software. Version numbers take the
form of \emph{MAJOR.MINOR{[}.PATCH{]}}, e.g., 0.2.6.
Changes in the major
version number herald significant changes in the software that are not
backwards compatible, such as changing or removing features or altering
the primary functions of the software. Increasing the minor version
represents incremental improvements in the software, like adding new
features. Following the minor version number can be an arbitrary number
of project-specific identifiers, including patches, builds and qualifiers.
Common qualifiers include \texttt{alpha}, \texttt{beta}, and \texttt{SNAPSHOT},
for applications that are
not yet stable or released, and \texttt{-RC} for release candidates prior
to an official release.

\ruleminor{The version of your software should be easily available by
supplying \texttt{-\/-version} or \texttt{-v} on the command line}. This command should print
the software name and version number, and it should
also be \ruleminor{included in all of the program's output}, particularly debugging
traces.  If someone needs help, it's important that they be able to tell
whoever's helping them which version of the software they're using.

While new releases may make a program better in general,
they can simultaneously create work for someone
who integrated the old version into their own workflow a year or two ago,
and won't see any benefits from upgrading.
A program's authors should therefore \ruleminor{ensure that old released versions
continue to be available.}
A number of mechanisms exist for
controlled release that range from as simple as adding an appropriate
commit message or tag to version control~\cite{blischak2016}, to official releases alongside
code on Bitbucket or GitHub, to depositing into a
repository like apt, yum, homebrew, CPAN, etc. Choose the method that
best suits the number and expertise of users you anticipate.

\rulemajor{Rule 5: Reuse software (within reason)}

In the spirit of code reuse and interoperability, developers often want
to reuse software written by others.
With a few lines, a call
is made out to \reviewed{another library or program} and the results are incorporated into the
primary script. Using popular projects reduces the amount of code that
needs to be maintained and leverages the work done by the other software.

Unfortunately, reusing software \reviewed{(whether software libraries or separate executables)}
introduces dependencies, which can
bring their own special pain. The interface between two software
packages can be a source of considerable frustration: all too
often, support requests descend into debugging errors produced by the
other project 
\reviewed{due to incompatible libraries, versions, or operating
systems~\cite{brown2013}. Even introducing libraries in the same programming
language can rely on software installed in the environment, and the problem
becomes much more difficult when relying on executables, or even on web
services.}

Despite these problems, software developers in research should
re-use existing software provided a few guidelines are adhered to.

First,
\ruleminor{make sure that you really need the auxiliary program}. If you are
executing GNU sort instead of figuring out how to sort lists in Python,
it may not be worth the pain of integration. \reviewed{Reuse software that offers some
measurable improvement to your project.}

Second, \reviewed{if launching an executable,}
\ruleminor{ensure the appropriate software and version is available}.
Either allow the user to
configure the exact path to the package, \reviewed{distribute the program with the
dependent software, or download it during installation using your
package manager. If the executable requires internet access, check for that
early in execution.

Third, \ruleminor{ensure that reused software is robust}. Relying on erratic third
party libraries or software is a recipe for tears. Prefer software that follows
good software development practices, is open for support questions, and is
available from a stable location or repository using your package manager.

Exercise caution especially when transitioning across languages or using
separate executables, as they tend to be especially sensitive to operating
systems, environments, and locales.}

\rulemajor{Rule 6: Rely on build tools and package managers for installation.}

\reviewed{To compile code, deploy applications, and automate other tasks,
programmers routinely use build tools like Make, Rake, Maven, Ant or MS Build.
These tools can also be used to manage runtime environments,
i.e.,
to check that the right versions of required packages are installed
and install or upgrade them if they are not.
As mentioned in Rule~5,
a package manager can mitigate some of the difficulties in software reuse.}

The same tools can and should be used to manage runtime environments on users' machines as well.
Accordingly,
developers should
\ruleminor{document all dependencies in a machine-readable form}.
Package managers like apt and yum are available on most Unix-like systems, and
application package managers exist for specific languages like Python (pip),
Java (Maven/Gradle), and Ruby (RubyGems). These package managers can be used
together with the build utility to ensure that dependencies are available at
compile/run time.

For example, it is common for Python projects to include a file called
\texttt{requirements.txt} that lists the names of required libraries,
along with version ranges:

\begin{verbatim}
requests>=2.0
pygithub>=1.26,<=1.27
python-social-auth>=0.2.19,<0.3
\end{verbatim}

This file can be read by the pip package manager, which can check that the
required software is available and install it if it is not.
Whatever is used,
developers should \emph{always} install dependencies
using their dependency description, especially on their personal machines, so that
they're sure it works.

Conversely, developers should
\ruleminor{avoid depending on scripts and tools which are not available as packages}.
In many cases, a program's author may not realize that some tool was built locally, and
doesn't exist elsewhere. At present, the only sure way to discover such
unknown dependencies is to install on a system administered by someone
else and see what breaks. As use of virtualization containers becomes more
widespread, software installation can also be tested on a virtual machine or
container system like Docker.

\rulemajor{Rule 7: Do not require root or other special privileges to install or run.}

Root (also known as ``superuser'' or ``admin'') is a special account on
a computer that has (among other things) the power to modify or delete
system files and user accounts. Conversely, files and directories owned
by root usually cannot be modifed by normal users.

Installing or running a program with root privileges is often
convenient, since doing so automatically bypasses all those pesky safety
checks that might otherwise get in the user's way. However, those checks
are there for a reason: scientific software packages may not
intentionally be malware, but one small bug or over-eager file-matching
expression can certainly make them behave as if they were. Outside of
very unusual circumstances,
\ruleminor{packages should not require root privileges to set up or use}.

Another reason for this rule is that users may want to try out a new
package before installing it system-wide on a cluster. Requiring root
privileges will frustrate such efforts, and thereby reduce uptake of the
package. Requiring, as Apache Tomcat does, that software be installed
under its own user account---i.e.,
that \texttt{packagename} be made a user, and all of the
package's software be installed in that pseudo-user's space---is similarly limiting,
and makes side-by-side installation of multiple versions of
the package more difficult.

Developers should therefore
\ruleminor{allow packages to be installed in an arbitrary location},
e.g., under a user's home directory in
\texttt{\textasciitilde{}/packagename}, or in directories with standard
names like \texttt{bin}, \texttt{lib}, and \texttt{man} under a chosen
directory. If the first option is chosen, the user may need to modify
her search path to include the package's executables and libraries, but
this can (more or less) be automated, and is much less risky than
setting things up as root.

Testing the ability to install software has traditionally been regarded as difficult,
since it necessarily alters the machine on which the test is conducted.
Lightweight virtualization containers like Docker make this much easier as well,
or simply \ruleminor{ask another person to try and build your software before releasing it}.

\rulemajor{Rule 8: Eliminate hard-coded paths.}

It's easy to write software that reads input from a file called
\texttt{mydata.csv}, but also very limiting. If a colleague asks you to
process her data, you must either overwrite your data file (which is
risky) or edit your code to read \texttt{otherdata.csv} (which is also
risky, because there's every likelihood you'll forget to change the
filename back, or will change three uses of the filename but not a
fourth).

Hard-coding file paths in a program also makes the software harder to run
in other environments. If your package is installed on a cluster, for
example, the user's data will almost certainly \emph{not} be in the same
directory as the software, and the folder
\texttt{C:\textbackslash{}users\textbackslash{}yourname\textbackslash{}}
will probably not even exist.

For these reasons, users should be able to
\ruleminor{set the names and locations of input and output files as command-line parameters}.
This rule applies to reference data sets as well as the user's own
data: if a user wants to try a new gene identification algorithm using
a different set of genes as a training set, she should not have to
edit the software to do so.
A corollary to this rule is
\ruleminor{do not require users to navigate to a particular directory to do their work},
since ``where I have to be'' is just another hard-coded path.

In order to save typing, it is often convenient to allow users to
specify an input or output \emph{directory}, and then require that there
be files with particular names in that directory. This practice, which
is sometimes called ``convention over configuration'', is used by many
software frameworks, such as WordPress and Ruby on Rails, and often
strikes a good balance between adaptability and consistency.

\rulemajor{Rule 9: Include a small test set that can be run to ensure the software is actually working.}

Every package should come with a set of tests for users to run
after installation. Its purpose is not only to check that the software
is working correctly (although that is extremely helpful), but also to
ensure that it works at all. This test script can also serve as a
working example of how to run the software.

In order to be useful, \ruleminor{make the tests easy to find and run}.
Many build systems will also run unit tests if provided them at compile time.
\reviewed{For users, or
if the build system is not amenable to testing, 
provide a working script
in the project's root directory named \texttt{runtests.sh}
or something equally obvious. 
This lets new users build their analysis from a working script.
For example, with its distribution, 
HISAT2 includes a full set of very small files,
and a 'Getting Started with HISAT2' section in its manual 
that leads you through the entire data lifecycle~\cite{pertea2016}.}

Equally, \ruleminor{make the test script's output easy to interpret}. Screens
full of correlation coefficients do not qualify: instead, the script's
output should be simple to understand for non-experts,
such as one line per test, with the test's name
and its pass/fail status, followed by a single summary line saying how
many tests were run and how many passed or failed. If many or all tests
fail because of missing dependencies, that fact should be displayed
once, clearly, rather than once per test, so that users have a clear
idea of what they need to fix and how much work it's likely to take.

Research has shown that the ease with which people can start making
contributions is a strong predictor of whether they will or not~\cite{steinmacher2015}.
By making it simpler for outsiders to contribute,
a test suite of any kind also makes it more likely that they will, and software
with collaborators stands a better chance of surviving in the busy field of
scientific software.

\rulemajor{Rule 10: Produce identical results when given identical inputs.}

\reviewed{The usage message tells users what the program could do.
It is equally important for the program to tell users what it actually did.
Accordingly,
when the program starts, it should \ruleminor{echo all parameters and software
versions to standard out or a log file alongside the results}. This
feature supports greater reproducibility because any result can be
replicated with only the previous output files as reference.}

Given a set of parameters and a dataset, \ruleminor{a particular version of a program
should produce the same results every time it is run}
to aid testing, debugging, and reproducibility.
Even minor changes to code can cause minor changes in output because of floating-point issues,
which means that getting exactly the same output for the same input and parameters
probably won't work during development,
but it should still be a goal for people who have deployed a specific version.

Many applications rely on randomized algorithms to
improve performance or runtimes. As a consequence, results can change
between runs, even when provided with the same data and parameters. By
its nature, this randomness renders strict reproducibility and therefore
debugging more difficult. If even the small test set (\#9) produces
different results for each run, new users may not be able to tell whether the software is
working properly. When comparing
results between versions or after changing parameters, even small
differences can confuse or muddy the comparison. And especially when
producing results for publications, grants or diagnoses, any analysis
should be absolutely reproducible.

Given the size of biological data, it is unreasonable to suggest that
random algorithms be removed. However, most programs use a pseudo-random
number generator, which uses a starting seed and an equation to
approximate random numbers. Setting the seed to a consistent value
can remove randomness between runs. \ruleminor{Allow the user to optionally provide
the random seed as an input parameter}, thus rendering the program deterministic
for those cases where it matters. If the seed is set internally (e.g.,
using clock time), echo it to the output for re-use later.
If setting the seed is not possible, \ruleminor{make sure the acceptable tolerance is
known and detailed in documentation and in the tests}.

\section*{Conclusion}

There has been extended discussion over the past few years of the
sustainability of research software, but this question is meaningless
in isolation: any piece of software can be sustained if its users are
willing to put in enough effort.  The real equation is the ratio
between the skill and effort available, and the ease with which
software can be installed, understood, used, maintained, and extended.
Following the ten rules we outline here reduce the denominator, and
thereby enable researchers to build on each other's work more easily.

That said, not \emph{every} coding effort needs to be engineered to
last.  Code that is used once to answer a specific question related to
a specific dataset doesn't require comprehensive documentation or
flexible configuration, and the only sensible way to test it may well
be to run it on the dataset in question. Exploratory analysis is an
iterative process that is developed quick and revised
often~\cite{lawlor2015,sanders2008}.  However, if a script is dusted
off and run three or four times for slightly different purposes, is
crucial to a publication or a lab, or being passed on to someone else,
it may be time to make your software more robust.

\end{document}